\documentclass[11pt]{article}

\usepackage[margin=1in]{geometry}
\usepackage[utf8]{inputenc}
\usepackage[T1]{fontenc}
\usepackage{amsmath,amssymb}
\usepackage{booktabs}
\usepackage{graphicx}
\usepackage{xcolor}
\usepackage{tikz}
\usetikzlibrary{arrows.meta,positioning,calc}
\usepackage[colorlinks=true,linkcolor=blue!60!black,citecolor=blue!60!black,urlcolor=blue!60!black]{hyperref}
\usepackage[round]{natbib}

\newcommand{\corpusA}{\textsc{Corpus~A}}

\title{Evidence-Based Scientific Question Discovery:\\
\large A Framework with Historical Backtesting}

\author{%
  Hui Mao\\
  Independent Researcher\\
  hui.mao@alumni.upenn.edu
}
\date{\today}

\begin{document}

\maketitle

\begin{abstract}
Current AI systems are optimized for answering questions; the scientific
enterprise is bottlenecked earlier, at \emph{discovering the questions
worth investigating}. We present a framework that turns a
traceable, reproducible, scope-controlled research corpus into ranked,
falsifiable research questions: evidence is represented as provenance-carrying
claims; cross-paper tensions are detected, typed, and human-adjudicated;
surviving signals are refined into questions and ranked by a two-stage
protocol separating scientific priority from execution priority. We
instantiate the framework on exoplanet atmospheres, a domain that uniquely
combines literature, structured catalogs, and space-telescope archives.
In a historical backtest, all questions generated from evidence available
before 2021 were substantively engaged by the 2021--2026 literature the
system never saw: two were answered---including one whose premise the
community later explicitly refuted---and the top-ranked question is
independently posed and still open. These results suggest that systematic
question discovery from evidence tensions surfaces the questions working
scientists subsequently invest in.
\end{abstract}

\section{Introduction}
\label{sec:introduction}

\begin{quote}
\emph{``What are the important problems of your field?\,\ldots\ If you do
not work on an important problem, it's unlikely you'll do important
work.''}\\
\hspace*{\fill}--- Richard Hamming, \emph{You and Your Research} (1986)
\end{quote}

Hamming's question is about asking, not answering. Current AI is
optimized for the opposite: benchmarks, training objectives, and
deployment all reward answering questions someone else has already
posed. Yet the scientific enterprise is bottlenecked earlier in the
loop---someone must notice that a question is worth asking. Recognizing
a productive question---an unexplained discrepancy, a conclusion resting
on a single dataset, a methodological assumption quietly doing
load-bearing work---is a distinct capability from answering it, and
arguably the rarer one. We take the position that this capability,
\emph{scientific question discovery}, is a core requirement for general
scientific intelligence, and that it can be studied rigorously today.

Studying question discovery requires discipline that question answering
does not. An answer can be checked against ground truth; a question's
value only reveals itself later, through the community's engagement with
it. This paper contributes a methodology that makes question discovery
measurable despite that asymmetry:

\begin{enumerate}
  \item \textbf{A framework} (Section~\ref{sec:framework}) that
    represents published evidence as provenance-carrying claims, detects
    and types cross-paper tensions in an evidence graph, refines
    surviving signals into falsifiable questions, and ranks them with a
    two-stage protocol that separates scientific priority from execution
    priority.
  \item \textbf{A corpus discipline} (Section~\ref{sec:benchmark}) in
    which every experiment is bound to a manifest that pins queries,
    versions, and---critically---a time boundary. Data need not live in
    the repository, but the queries, versions, time boundaries, and
    processing code that produce it must.
  \item \textbf{A historical validation protocol}
    (Section~\ref{sec:benchmark}) that exploits the time boundary:
    questions are generated from evidence available before a cutoff and
    judged against the literature published after it, which the system
    never saw. This prevents the system from peeking at the future and
    converts question value into a measurable outcome.
  \item \textbf{An end-to-end instantiation}
    (Section~\ref{sec:experiments}) on exoplanet atmospheres, chosen
    because it uniquely combines papers, structured catalogs (NASA
    Exoplanet Archive), and space-telescope archives (JWST/HST via
    MAST), allowing the full loop---evidence, questions, falsification,
    ranking---to run end to end.
\end{enumerate}

In the historical backtest, every question generated from
pre-2021 evidence was substantively engaged by the 2021--2026 literature;
two were answered, including one whose premise the community later
explicitly refuted---the strongest form of question-discovery success,
since the system had isolated a published conclusion that was in fact an
artifact awaiting correction.

\section{Related Work}
\label{sec:related}

\paragraph{Automated scientific discovery.}
Computational discovery has a long lineage: \textsc{Bacon} and its
successors rediscovered empirical laws from data
\citep{langley1987scientific}, and the Robot Scientist line (Adam, Eve)
closed the loop from hypothesis generation through physical
experimentation in functional genomics
\citep{king2004robot,king2009automation}. Contemporary
``agentic-scientist'' systems extend this ambition with large language
models: the AI Scientist generates ideas and drafts papers end to end
\citep{lu2024aiscientist}, Google's AI co-scientist generates and
tournament-ranks research hypotheses with multi-agent debate
\citep{gottweis2025coscientist}, and deep-research agents synthesize
literature into reports \citep{openai2025deepresearch}. Across this
line, the object produced is a hypothesis, an experiment, or a paper;
the selection of \emph{which question deserves the loop} is either
inherited from the human operator or made implicitly inside a prompt.
Our system makes that selection step itself the object of study, with
its own representations, human checkpoints, and---critically---an
outcome-based evaluation.

\paragraph{Literature understanding and scientific knowledge bases.}
Our evidence layer builds on infrastructure for machine-readable
science: literature graphs such as Semantic Scholar
\citep{ammar2018construction} and OpenAlex \citep{priem2022openalex},
structured claim resources like the Open Research Knowledge Graph
\citep{jaradeh2019orkg}, scientific claim verification
\citep{wadden2020scifact}, science-specialized language models
\citep{taylor2022galactica}, and retrieval-augmented literature agents
\citep{lala2023paperqa,skarlinski2024superhuman}. We differ in what the
representation is \emph{for}: claims carry verification tiers and
source locations, and relations between claims are typed tension edges
(\emph{contradicts, qualifies, challenges-method, explains-discrepancy,
complements, supports}) adjudicated by a human---an empirical necessity,
since our review found only 1 of 16 top tension candidates to be a
genuine contradiction (Section~\ref{sec:experiments:tension}).

\paragraph{Open-ended learning and intrinsic motivation.}
Curiosity-driven agents formalize ``interestingness'' via intrinsic
motivation \citep{oudeyer2007intrinsic,schmidhuber2010formal}, novelty
search \citep{lehman2011abandoning}, and the broader open-endedness
program \citep{stanley2017openendedness}. Those signals are computed
from an agent's own experience stream; ours are computed from the
evidential structure of the scientific record---contradiction,
single-source support, unexamined assumptions---which is where working
scientists actually look for questions.

\paragraph{LLM research ideation and its evaluation.}
Recent studies evaluate LLM-generated research ideas with expert panels,
finding above-human novelty ratings alongside feasibility concerns
\citep{si2024canllms}; LLM-as-judge protocols and their biases are
themselves under scrutiny \citep{zheng2023judging}. Panel evaluation
asks contemporaries to \emph{predict} a question's value. Our historical
backtest removes that prediction: questions are generated from a
time-frozen corpus and scored by what the field actually did in the
following five years.

\begin{table}[t]
  \centering
  \caption{Positioning relative to representative systems. ``Evidence
  tensions'' = questions/hypotheses grounded in explicit cross-paper
  evidential structure; ``historical validation'' = outcome-based
  evaluation against literature published after a data cutoff.}
  \label{tab:comparison}
  \small
  \begin{tabular}{lcccc}
    \toprule
    System & Hypotheses & Questions & Tensions & Backtest \\
    \midrule
    \textsc{Bacon} \citep{langley1987scientific} & \checkmark & -- & -- & -- \\
    Robot Scientist \citep{king2009automation} & \checkmark & -- & -- & -- \\
    AI Scientist \citep{lu2024aiscientist} & \checkmark & -- & -- & -- \\
    AI co-scientist \citep{gottweis2025coscientist} & \checkmark & -- & -- & -- \\
    PaperQA2 \citep{skarlinski2024superhuman} & -- & -- & -- & -- \\
    LLM ideation \citep{si2024canllms} & \checkmark & \checkmark & -- & -- \\
    \midrule
    This work & -- & \checkmark & \checkmark & \checkmark \\
    \bottomrule
  \end{tabular}
\end{table}
\section{Framework}
\label{sec:framework}

\begin{figure}[t]
  \centering
  \begin{tikzpicture}[
    font=\small,
    box/.style={rectangle, rounded corners=2pt, draw=black!55, align=center,
                minimum width=3.6cm, minimum height=0.66cm, inner sep=3pt},
    pipe/.style={box, fill=blue!10},
    human/.style={box, fill=orange!18},
    arr/.style={-{Stealth[length=2.2mm]}, semithick, black!65},
    duo/.style={{Stealth[length=2.2mm]}-{Stealth[length=2.2mm]}, semithick, black!65},
    node distance=0.34cm
  ]
    \node[pipe] (lit) {Scientific Literature};
    \node[pipe, below=of lit] (claims) {Claims};
    \node[pipe, below=of claims] (graph) {Evidence Graph};
    \node[pipe, below=of graph] (tension) {Typed Tensions};
    \node[pipe, below=of tension] (fam) {Question Families};
    \node[pipe, below=of fam] (refine) {Question Refinement};
    \node[pipe, below=of refine] (rank) {Two-stage Ranking};
    \node[pipe, below=of rank] (back) {Historical Backtesting};

    \node[human, right=2.2cm of graph] (review) {Human Review};
    \node[human, right=2.2cm of tension] (adjud) {Typed Tension\\Adjudication};
    \node[human, right=2.2cm of refine] (curate) {Question Curation};
    \node[human, right=2.2cm of back] (valid) {Historical Validation};

    \foreach \a/\b in {lit/claims, claims/graph, graph/tension,
                       tension/fam, fam/refine, refine/rank, rank/back}
      \draw[arr] (\a) -- (\b);
    \draw[arr] (review) -- (adjud);
    \draw[arr] (adjud) -- (curate);
    \draw[arr] (curate) -- (valid);

    \draw[duo] (tension) -- (adjud);
    \draw[duo] (refine) -- (curate);
    \draw[duo] (back) -- (valid);
    \draw[arr] (review.west) -- (graph.east);
    \draw[arr, dashed] (valid.east) to[bend right=70, looseness=1.4]
      node[right=3pt, pos=0.5, align=left, font=\scriptsize\itshape]
      {verdicts become\\graph edges}
      (review.east);
  \end{tikzpicture}
  \caption{Architecture: a dual loop. The automated pipeline (left,
  blue) turns a time-frozen corpus into ranked questions; the human loop
  (right, amber) adjudicates tension types, curates question phrasing,
  and validates outcomes. Human verdicts are committed as data and feed
  back into the evidence graph, keeping expert judgment cheap,
  auditable, and reusable.}
  \label{fig:architecture}
\end{figure}
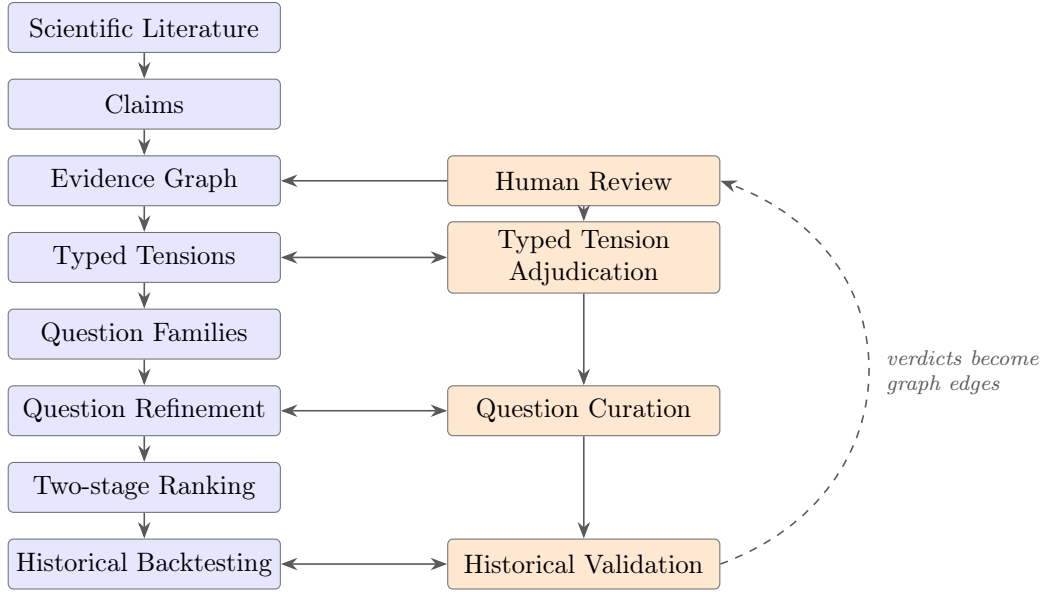

The framework transforms existing scientific knowledge into ranked,
falsifiable questions through six stages
(Figure~\ref{fig:architecture}). Throughout, the design principle is
\emph{traceability}: every artifact carries the provenance needed to
audit or reproduce it.

\subsection{Evidence representation}
\label{sec:framework:evidence}

Knowledge enters as three layers, collected in order of increasing cost:
a \emph{literature} layer (what has been claimed), a \emph{catalog}
layer (what objects exist, with parameters and uncertainties), and an
\emph{observation} layer (where the underlying data lives---metadata
only; data products are fetched on demand after a question is selected).
Every raw record is stored immutably inside a provenance envelope
$(\textit{source}, \textit{retrieved-at}, \textit{query-version},
\textit{source-id}, \textit{payload})$, then normalized into relational
tables.

The atomic unit is the \textbf{claim}: a tuple
\[
c = (\textit{text},\ \textit{evidence},\ \textit{assumptions},\
     \textit{uncertainty},\ \textit{objects},\ \textit{datasets},\
     \textit{location},\ \textit{tier})
\]
where \textit{location} points back into the source document and
\textit{tier} records the verification provenance:
\texttt{human\_gold} (annotated by a domain reviewer),
\texttt{model\_matched\_gold} (model extraction semantically matching a
gold claim; the canonical text remains the human's, both provenances are
kept), or \texttt{model\_only} (unreviewed model extraction, default
confidence 0.75). Model claims that duplicate gold claims are absorbed
into a single canonical node rather than added as siblings---duplicate
nodes would both inflate the graph and manufacture spurious tension.

\subsection{Evidence graph and typed tensions}
\label{sec:framework:graph}

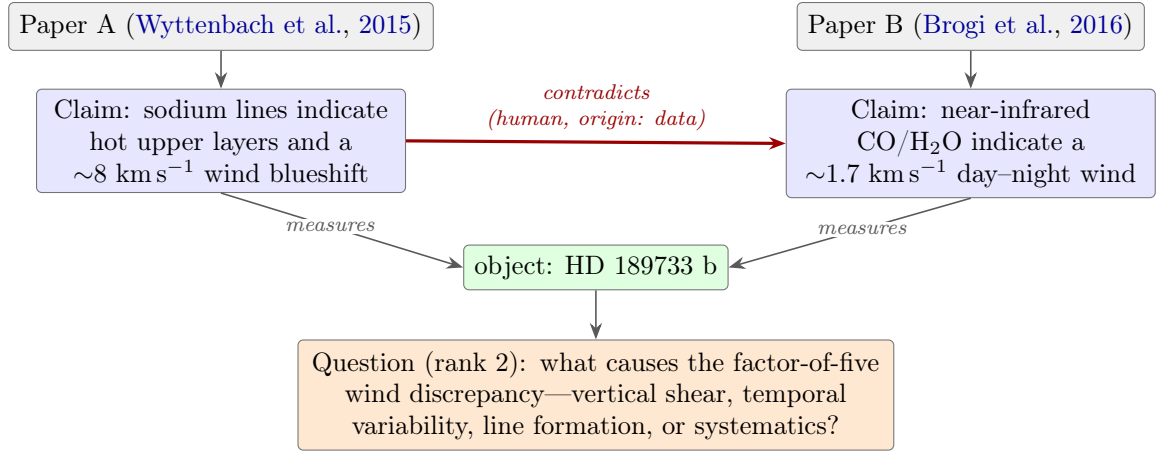
\begin{figure}[t]
  \centering
  \begin{tikzpicture}[
    font=\small,
    box/.style={rectangle, rounded corners=2pt, draw=black!55, align=center,
                inner sep=4pt},
    paper/.style={box, fill=black!6, minimum width=3.4cm},
    claim/.style={box, fill=blue!10, text width=4.6cm},
    entity/.style={box, fill=green!12},
    question/.style={box, fill=orange!18, text width=7.6cm},
    arr/.style={-{Stealth[length=2.2mm]}, semithick, black!65},
    lab/.style={font=\scriptsize\itshape, fill=white, inner sep=1pt}
  ]
    \node[paper] (pA) {Paper A \citep{wyttenbach2015}};
    \node[paper, right=4.8cm of pA] (pB) {Paper B \citep{brogi2016}};
    \node[claim, below=0.5cm of pA] (cA)
      {Claim: sodium lines indicate hot upper layers and a
       ${\sim}8$~km\,s$^{-1}$ wind blueshift};
    \node[claim, below=0.5cm of pB] (cB)
      {Claim: near-infrared CO/H$_2$O indicate a
       ${\sim}1.7$~km\,s$^{-1}$ day--night wind};
    \node[entity] (obj) at ($(cA.south)!0.5!(cB.south) + (0,-0.95)$)
      {object: HD~189733~b};
    \node[question, below=0.65cm of obj] (q)
      {Question (rank 2): what causes the factor-of-five wind
       discrepancy---vertical shear, temporal variability,
       line formation, or systematics?};

    \draw[arr] (pA) -- (cA);
    \draw[arr] (pB) -- (cB);
    \draw[arr] (cA.south) -- node[lab, pos=0.45] {measures} (obj.west);
    \draw[arr] (cB.south) -- node[lab, pos=0.45] {measures} (obj.east);
    \draw[arr, very thick, red!60!black] (cA.east) --
      node[lab, above=3pt, align=center, text=red!60!black]
      {contradicts\\(human, origin: data)} (cB.west);
    \draw[arr] (obj) -- (q);
  \end{tikzpicture}
  \caption{The evidence graph around the one confirmed contradiction:
  two claims from different instruments measure the same object,
  disagree on a shared quantity, survive human adjudication as a
  genuine data-origin tension, and become the pipeline's top
  tension-driven question. The 15 other reviewed candidates received
  non-contradiction types and correspondingly weaker question
  priority.}
  \label{fig:graph-example}
\end{figure}

Claims, objects, datasets, and assumptions become nodes; deterministic
rules add edges (\textit{uses-dataset}, \textit{measures},
\textit{assumes}), each recording the rule that created it. Cross-paper
claim pairs that share an object \emph{and} a topic term become
\emph{tension candidates}---deliberately a weak, high-recall rule. The
key representational commitment is that candidate tensions are
\textbf{typed by human adjudication} before they may drive question
generation, using the vocabulary
\[
\{\textit{contradicts},\ \textit{qualifies},\
  \textit{challenges-method},\ \textit{explains-discrepancy},\
  \textit{complements},\ \textit{supports}\}.
\]
This vocabulary is an empirical result of the calibration review
(Section~\ref{sec:experiments:tension}): most high-scoring candidates
are not contradictions, and a graph that collapses ``adds limiting
conditions'' into ``negates'' will amplify false conflicts as it grows.
Figure~\ref{fig:graph-example} shows the instantiation of this
machinery on the one confirmed contradiction in our corpus.

\subsection{Question generation}
\label{sec:framework:questions}

Questions are generated only from prioritized signal classes:
(P1) human-confirmed observational tensions, (P2) methodological
challenges, (P3) qualifications across independent datasets, and
(P4) single-dataset conclusions from trusted claims. Unreviewed tension
candidates are excluded by design. Signal assembly is deterministic;
a language model only phrases the question, and every candidate carries
its evidence trail (claim identifiers, papers, adjudication basis) plus
an archival-data availability count---the first falsification screen.
Human editorial decisions (merges into question families, rewrites,
rejections) are recorded as committed curation records and re-applied
deterministically.

\subsection{Two-stage ranking}
\label{sec:framework:ranking}

Stage~A is a quality gate: evidence grounding plus a six-criterion
clarity rubric (presupposed conclusions, statistics/physics conflation,
explicit comparison, falsifiable outcome, bounded scope, stateable
failure condition). Clarity gates rather than scores---in calibration it
scored uniformly and only added a constant. Human-curated questions
bypass the model gate, whose verdict is retained as advisory.

Stage~B ranks survivors by
\[
\textit{scientific priority} =
0.35\,S + 0.25\,T + 0.20\,F + 0.10\,N + 0.10\,G,
\]
where $S$ (significance) is judged by a model but hard-capped by signal
tier (P1:~10, P2:~9, P3:~8.5, P4:~7.5) so that single-dataset robustness
checks cannot rank beside confirmed tensions; $T$ (tension strength) is
structural; $F$ (feasibility) averages judged subcomponents
(availability of the \emph{right} data rather than raw record counts,
data independence, analysis readiness, reanalysis sufficiency);
$N$ (novelty) is deliberately down-weighted---the system generates
questions from existing evidence tensions, so surface novelty is not the
point---and blends a judge with a corpus-similarity check; $G$ is
expected information gain. \emph{Execution priority} (feasibility) is
reported separately: abundant archives must not outrank a confirmed
cross-instrument tension, and a question can be scientifically first
while operationally harder.

\section{Corpus Discipline and Evaluation Protocol}
\label{sec:benchmark}

\subsection{Corpus manifests}

Every experiment must know exactly what the system has seen. A
\emph{corpus manifest} pins the scope, source queries, per-source field
lists, record caps, selection rules, and---critically---a
\texttt{cutoff\_date}. Manifests are frozen at collection time and
referenced by experiments; a change of scope mints a new corpus
identifier rather than editing a manifest in place. The repository
commits manifests, collectors, schemas, curation and adjudication
records, and small samples; bulk data is regenerable from those.

\subsection{Historical validation}
\label{sec:benchmark:validation}

The cutoff enables a backtest that removes guesswork about the future
from evaluation:

\begin{enumerate}
  \item \corpusA{} contains only evidence available up to the cutoff
    (here 2020-12-31), enforced end to end: literature windows, catalog
    snapshots, and a contamination guard that filters every normalized
    table by corpus identifier.
  \item The full pipeline runs on \corpusA{} to produce ranked
    questions.
  \item A \emph{validation corpus} of post-cutoff literature
    (2021--2026), never seen by the pipeline, is collected under its own
    manifest. For each ranked question, the most relevant post-cutoff
    abstracts are retrieved by embedding similarity and a judge
    classifies the question's fate---\texttt{answered},
    \texttt{partially\_addressed}, \texttt{posed\_but\_open}, or
    \texttt{not\_addressed}---citing only retrieved records, and states
    whether the question's premise was validated, refuted, or left
    untested.
\end{enumerate}

\begin{figure}[t]
  \centering
  \begin{tikzpicture}[
    font=\small,
    box/.style={rectangle, rounded corners=2pt, draw=black!55, align=center,
                minimum height=0.66cm, inner sep=4pt},
    past/.style={box, fill=blue!10},
    future/.style={box, fill=green!12},
    res/.style={box, fill=orange!18},
    arr/.style={-{Stealth[length=2.2mm]}, semithick, black!65}
  ]
    \draw[very thick, black!50, -{Stealth[length=3mm]}] (0,0) -- (13.6,0);
    \draw[thick, red!60!black, dashed] (5.4,-0.55) -- (5.4,2.35)
      node[above, font=\scriptsize\itshape, text=red!60!black, align=center]
      {cutoff\\2020-12-31};
    \node[font=\scriptsize, black!60] at (2.6,-0.35) {2015--2020};
    \node[font=\scriptsize, black!60] at (9.4,-0.35) {2021--2026};

    \node[past]   (corpus) at (1.6,1.0) {\corpusA};
    \node[past]   (gen)    at (4.2,1.0) {Generate\\\& Freeze};
    \node[future] (vcorp)  at (7.0,1.0) {Validation\\Corpus};
    \node[future] (retr)   at (9.3,1.0) {Retrieve};
    \node[future] (judge)  at (11.2,1.0) {Judge};
    \node[res]    (outc)   at (13.2,1.0) {Outcome};

    \draw[arr] (corpus) -- (gen);
    \draw[arr] (gen) -- (vcorp);
    \draw[arr] (vcorp) -- (retr);
    \draw[arr] (retr) -- (judge);
    \draw[arr] (judge) -- (outc);

    \node[font=\scriptsize, align=left, black!70] at (13.0,2.0)
      {answered: 2\\posed, open: 1\\partial: 7\\ignored: 0};
  \end{tikzpicture}
  \caption{Historical backtesting. Questions are generated from a
  corpus frozen at the cutoff, then judged against post-cutoff
  literature the pipeline never saw. The right panel shows the observed
  outcome distribution for the ten ranked questions
  (Section~\ref{sec:experiments:validation}).}
  \label{fig:backtest}
\end{figure}
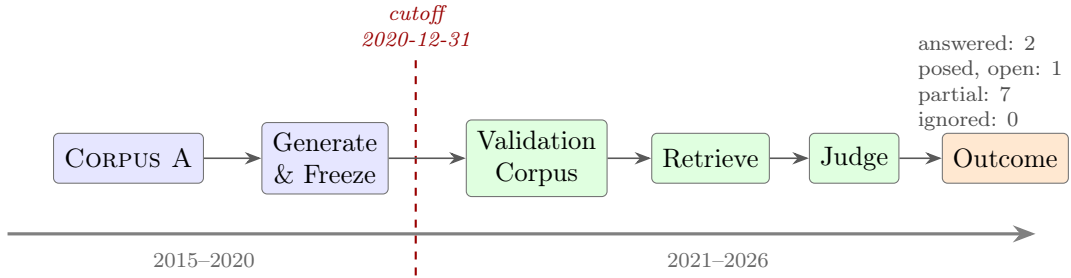

A question the community independently posed, answered, or refuted after
the cutoff is evidence the pipeline surfaces questions worth asking; a
premise the community later refuted is the strongest success mode, since
the system isolated a published conclusion that was an artifact awaiting
correction. Figure~\ref{fig:backtest} summarizes the protocol.

\subsection{Question quality rubric}

Human evaluation shares one rubric with the ranking stage: novelty,
feasibility, significance, and clarity, with clarity applied as a gate
(Section~\ref{sec:framework:ranking}). All rubric prompts, gate
criteria, weights, and per-component scores are committed, making every
ranking auditable.

\section{Experiments}
\label{sec:experiments}

We instantiate the framework on exoplanet atmospheres---specifically,
atmospheric composition interpretation in transmission spectra,
cloud/haze degeneracies, and instrument systematics. All numbers below
are produced by the committed pipeline under the frozen manifest
\texttt{exoplanet\_atmospheres\_v1} (cutoff 2020-12-31).

\subsection{Corpus construction}
\label{sec:experiments:corpus}

Table~\ref{tab:corpus} summarizes \corpusA{}. Literature metadata came
from NASA ADS (primary) and arXiv (supplement) under four topic queries;
cross-source deduplication by arXiv identifier removed 404 duplicate
records. A core set of 500 papers was selected by greedy,
year-stratified optimization over internal citation in-degree, keyword
coverage, and target diversity (83--84 papers per year, 2015--2020).
Object mentions were linked to the NASA Exoplanet Archive by
boundary-anchored name matching tolerant of space/hyphen variants;
naive substring matching produced systematic false positives
(``WASP-1'' firing inside ``WASP-12b'') that inflated links by
$2.3\times$ before the fix.

\begin{table}[t]
  \centering
  \caption{\corpusA{} at a glance (manifest
  \texttt{exoplanet\_atmospheres\_v1}, cutoff 2020-12-31).}
  \label{tab:corpus}
  \begin{tabular}{lr}
    \toprule
    Literature metadata (deduplicated) & 2{,}512 papers \\
    \quad from NASA ADS / arXiv & 1{,}639 / 1{,}277 \\
    Core full-text set & 500 papers \\
    Citation edges & 51{,}616 \\
    Catalog objects (NASA Exoplanet Archive) & 6{,}324 planets \\
    Observation metadata (MAST, JWST+HST spectra) & 306{,}729 records \\
    Paper--object links & 2{,}364 \\
    Catalog planets with archival spectra & 553 \\
    \bottomrule
  \end{tabular}
\end{table}

\subsection{Claim extraction: calibration before scale}
\label{sec:experiments:claims}

A domain reviewer first annotated 20 claim-dense core papers by hand,
yielding 37 gold claims with source locations, assumptions, and
uncertainties. Model extraction (from abstracts in this version) was
then calibrated against the gold set over three prompt revisions:
the initial prompt over-split coupled measurements, occasionally copied
first-person phrasing, and preferred easily-copied quantitative results
over interpretive conclusions (e.g., it captured ``mostly subsolar
H$_2$O'' across ten hot Jupiters \citep{pinhas2019} but dropped the
paper's load-bearing interpretation of subsolar oxygen or supersolar
C/O). The final prompt enforces 2--5 load-bearing claims with a priority
order (conclusion $>$ interpretation $>$ methodological challenge $>$
decisive quantity $>$ secondary measurement), merging of coupled
measurements, normalized third person, and an omission self-check.
Under it, all reviewer-cited misses were recovered, first-person copying
fell to zero, and no paper exceeded five claims (mean 3.68 across
353 claims from 100 papers; 4 methods papers correctly yielded none).

Semantic deduplication against the gold set (same paper, compatible
objects, same conclusion polarity, embedding cosine $\ge 0.62$) matched
all 37 gold claims to a model counterpart (similarity 0.63--0.97, median
0.76), producing a canonical table of 353 claims: 37
\texttt{model\_matched\_gold} and 316 \texttt{model\_only}.

\subsection{Tension detection and adjudication}
\label{sec:experiments:tension}

The high-recall candidate rule produced 161 cross-paper tension
candidates over the canonical claims (807-node, 1{,}237-edge graph).
The reviewer adjudicated the top-ranked candidates; the resulting
16 typed verdicts are themselves a finding
(Table~\ref{tab:adjudication}): only \emph{one} candidate is a genuine
observational contradiction---the HD~189733~b wind-velocity discrepancy,
${\sim}8$~km\,s$^{-1}$ from optical sodium \citep{wyttenbach2015} versus
${\sim}1.7$~km\,s$^{-1}$ from near-infrared CO/H$_2$O
\citep{brogi2016}. The modal relations are \textit{supports} and
\textit{complements} (e.g., terminator transmission versus dayside
emission water results for HD~209458~b probe different geometries and do
not conflict), followed by \textit{challenges-method} (a 1D-retrieval
bias result \citep{macdonald2020} limits the interpretation of a water
abundance \citep{kreidberg2015} without negating the detection). An
early binary confirm/reject protocol would have discarded most of these
relationships or, worse, upgraded them to contradictions.

\begin{table}[t]
  \centering
  \caption{Human adjudication of the 16 reviewed tension candidates.}
  \label{tab:adjudication}
  \begin{tabular}{lr}
    \toprule
    supports & 7 \\
    complements & 4 \\
    challenges-method & 2 \\
    qualifies & 1 \\
    explains-discrepancy & 1 \\
    contradicts (confirmed observational tension) & 1 \\
    \bottomrule
  \end{tabular}
\end{table}

\subsection{Question generation and two-stage ranking}
\label{sec:experiments:ranking}

From the prioritized signals (1~confirmed tension, 2~methodological
challenges, 1~independent qualification, 7~trusted single-dataset
conclusions), the generator produced 11 candidates; editorial curation
merged one near-duplicate pair into a question family with two
sub-questions and rewrote three questions (one had conflated statistical
significance with physical conditions; one presupposed the tension's
favored explanation; one was reframed for historical validation). The
Stage-A gate additionally caught the presupposition defect
independently. After curation, all 10 questions pass the gate.

Stage-B ranking places the terminator-heterogeneity family and the
wind-discrepancy question statistically tied at the top (scientific
priority 8.13 vs.\ 8.13, separated by 0.005) with opposite execution
profiles: the heterogeneity family is reanalysis-ready against 2{,}643
archival HST spectra of WASP-12 (execution priority 9.1), whereas the
confirmed tension is scientifically first-equal but operationally
harder (execution 6.9---the needed simultaneous multi-band data may not
exist). Significance caps hold the seven P4 robustness questions to a
6.6--7.1 band, below both tension-driven questions.

\subsection{Historical validation}
\label{sec:experiments:validation}

The validation corpus contains 1{,}891 unique 2021--2026 papers
(collected under its own manifest; a contamination guard keeps
\corpusA{} tables bit-identical before and after collection).
Table~\ref{tab:validation} lists all ten questions with their verdicts:
2 answered, 1 posed but open, 7 partially addressed, 0 not addressed.

\begin{table}[t]
  \centering
  \caption{Historical validation of all ten ranked questions (generated
  from pre-2021 evidence, judged against 2021--2026 literature).
  Premise: whether the post-cutoff literature validated, refuted, or
  left untested the evidential premise behind the question.}
  \label{tab:validation}
  \small
  \begin{tabular}{clp{7.6cm}ll}
    \toprule
    Rank & ID & Question (abridged) & Verdict & Premise \\
    \midrule
    1 & q\_002 & WASP-12\,b: terminator heterogeneity vs.\ water
        evidence, abundance, C/O & posed, open & untested \\
    2 & q\_001 & HD~189733\,b: cause of factor-of-five wind
        discrepancy (Na vs.\ CO/H$_2$O) & partial & validated \\
    3 & q\_004 & HD~189733\,b: conditions for subsolar H$_2$O with
        robust CO & partial & validated \\
    4 & q\_008 & HD~209458\,b: subsolar water---atmospheric reality or
        retrieval artifact? & \textbf{answered} & \textbf{refuted} \\
    5 & q\_007 & Patchy-cloud solar-composition models under
        independent retrievals \& data & partial & untested \\
    6 & q\_010 & WASP-121\,b: water detection sensitivity to clouds,
        contamination, priors, null model & partial & untested \\
    7 & q\_005 & HD~209458\,b: independent support for NH$_3$/HCN
        across cloud scenarios & \textbf{answered} & validated \\
    8 & q\_009 & Conditions for patchy-cloud fits without high mean
        molecular weight & partial & validated \\
    9 & q\_006 & WASP-12\,b: does C/O~$>$~1 exclusion survive without
        equilibrium chemistry? & partial & untested \\
    10 & q\_011 & TRAPPIST-1: have JWST observations validated
        pre-launch CO$_2$ detectability predictions? & partial & untested \\
    \bottomrule
  \end{tabular}
\end{table}

Three outcomes deserve emphasis. First, \textbf{a premise refuted}: the
question of whether HD~209458~b's strongly subsolar terminator water
abundance \citep{macdonald2017} is atmospheric reality or retrieval
artifact was answered by 2025 reanalyses using independent retrieval
frameworks with improved systematics treatment, which find
solar-consistent water---the pipeline had isolated a published
conclusion that was in fact an artifact awaiting correction. Second,
\textbf{the top question is independently posed and open}: the
terminator-heterogeneity family is actively pursued by the community
without resolution. Third, \textbf{the reframed historical-validation
case behaves as designed}: JWST/NIRSpec PRISM programs are testing the
pre-launch TRAPPIST-1 CO$_2$ detectability prediction
\citep{lustigyaeger2019}, with stellar contamination \citep{rackham2018}
currently blocking a decisive verdict---the judge correctly reports the
premise as untested rather than refuted.

That zero questions were ignored by the subsequent literature is the
central quantitative result: questions surfaced systematically from
evidence tensions in pre-2021 knowledge are the questions the field
subsequently invested in.

\subsection{Failure analysis}
\label{sec:experiments:failures}

The seven \emph{partial} verdicts are not one failure mode but three,
and each points at a specific pipeline improvement.

\paragraph{Blocked by a shared upstream obstacle.}
For the TRAPPIST-1 prediction test (rank 10) and the M-dwarf-adjacent
cloud questions (rank 5), the post-cutoff literature engaged exactly the
right test but could not reach a verdict because \emph{stellar
contamination} limits the achievable precision. Notably, the pipeline's
own corpus contained the warning \citep{rackham2018}: had the
falsification screen consulted the evidence graph for known
systematics on the target's host star, these questions could have been
generated with the obstacle named and a contamination-robust test
demanded. Failure source: the falsification stage checks data
\emph{availability} but not data \emph{sufficiency}.

\paragraph{Questions demanding joint analyses no single study performs.}
The wind-discrepancy question (rank 2) needs contemporaneous
optical--near-infrared spectroscopy; the subsolar-H$_2$O-with-CO
question (rank 3) needs both species constrained in one dataset. The
community advanced each piece separately. These questions are well
posed but implicitly assume a coordinated analysis; an execution-aware
refinement step could split them into staged sub-questions the way the
curated rank-1 family already is.

\paragraph{Questions broader than any single study answers.}
The class-level conditions question (rank 8) and the five-factor
sensitivity question (rank 6) span parameter spaces the literature
answers instance by instance. The clarity gate checks that scope is
\emph{bounded}, not that it matches the granularity at which the field
publishes; a granularity criterion would have flagged both.

\paragraph{Why zero \emph{not-addressed}.}
Beyond genuine question quality, two selection effects contribute:
tensions among highly-cited claims concern targets the community
already observes intensively, and the validation corpus was collected
with target-specific queries that guarantee topical coverage. The
baseline experiments proposed in Section~\ref{sec:discussion} (random
future-work questions; graph-free LLM questions over the same corpus)
are designed to measure exactly how much of the engagement rate these
effects explain.

\section{Discussion}
\label{sec:discussion}

\paragraph{What the backtest does and does not show.}
Historical validation measures whether generated questions align with
where the field actually went---strong evidence of question quality, but
conservative in one direction: a question the community never engaged
might still be valuable (the field can be collectively wrong or
resource-constrained). Conversely, engagement is partially confounded by
obviousness; a question everyone was about to ask scores well. The
premise-refuted outcome is the cleanest signal, since it requires the
system to have flagged a specific published conclusion as fragile before
the community overturned it. Scaling from 10 questions to hundreds, and
comparing against baselines (questions sampled from review-paper
future-work sections; questions generated without the evidence graph),
is the necessary next step before strong claims.

\paragraph{Human adjudication is load-bearing---by design.}
Tension typing and question curation are human steps, and the reviewed
sample shows why: only 1 of 16 top candidates was a true contradiction.
The framework's contribution is to make expert judgment cheap and
auditable (typed verdicts over ranked candidates, committed as data)
rather than to eliminate it. Automating adjudication against the
growing corpus of human verdicts---with the 33\% early-precision figure
as the baseline to beat---is a measurable follow-on task.

\subsection{Threats to validity}
\label{sec:discussion:threats}

\begin{description}
  \item[Single domain (astronomy only).] Exoplanet atmospheres was
    chosen because literature, catalogs, and observation archives align
    unusually well. The framework's interfaces are domain-agnostic, but
    the tension topic vocabulary and falsification screens are
    domain-supplied; the engagement rate may not transfer to fields with
    looser evidential structure. Porting to a second domain is the
    direct test.
  \item[Abstract-only extraction.] Claims are extracted from abstracts.
    Calibration showed abstracts carry the load-bearing conclusions, but
    assumptions and uncertainty statements often live in the body;
    full-text ingestion (with the chunk-level source locations the
    schema already mandates) should raise claim coverage and
    adjudication quality.
  \item[Single human adjudicator.] All tension verdicts and curation
    decisions come from one domain reviewer. Verdicts are committed
    verbatim and re-appliable, so inter-annotator agreement can be
    measured by replaying the same candidates past additional
    reviewers---but it has not yet been.
  \item[Judge bias.] The validation judge is an LLM; engagement
    judgments (``substantive'' vs.\ topical) retain subjectivity even
    with citations restricted to retrieved records. The extraction,
    ranking, and validation stages also share a model family, risking
    correlated blind spots. Mitigations in place: forced bibcode-grounded
    evidence, human spot-checks of verdicts; mitigations still needed:
    independent judges, human re-scoring of a verdict sample.
  \item[Small sample, no baseline.] Ten questions support existence
    claims, not rates. The engagement-rate confounds identified in the
    failure analysis (highly-cited targets, targeted validation queries)
    require the baseline comparisons described above before the headline
    numbers can be read as effect sizes.
\end{description}

\paragraph{Implications for general scientific intelligence.}
The results support a specific, testable position: question-asking
competence can be decomposed into auditable stages---evidence
representation, tension detection, refinement, prioritization---each of
which can be measured and improved independently, with historical
backtesting as the end-to-end score. If artificial general intelligence
is to include the scientist's core skill of knowing \emph{what to ask},
benchmarks of this form---rather than open-ended ideation judged by
contemporaneous panels---are the right instrument for tracking progress
toward it.
\section{Conclusion}
\label{sec:conclusion}

We presented a framework that treats scientific question discovery as a
first-class, measurable capability: evidence becomes provenance-carrying
claims; cross-paper tensions are detected and typed under human
adjudication; prioritized signals become falsifiable questions; and a
two-stage protocol ranks scientific priority separately from execution
priority. A corpus discipline built on frozen, time-bounded manifests
makes the whole loop reproducible and enables historical backtesting.

Instantiated on exoplanet atmospheres, the pipeline generated 10
questions from evidence available before 2021. The 2021--2026
literature---never seen by the system---engaged every one of them:
two were answered, one with its premise explicitly refuted; the
top-ranked question is independently posed and remains open. Question
discovery, these results suggest, is not an ineffable spark but a
capability that can be engineered, audited, and scored---and therefore
one on which progress toward general scientific intelligence can be
tracked.

\bibliographystyle{plainnat}
\bibliography{references}

\end{document}